\documentclass[aps,amsfonts,nofootinbib,superscriptaddress,twocolumn]{revtex4-1}

\usepackage{appendix}
\usepackage{dsfont}
\usepackage{amssymb}
\usepackage{graphicx}
\usepackage{amsmath}
\usepackage{amsbsy}
\usepackage{color}
\usepackage{slashed}
\usepackage{bm}
\usepackage{bbm}
\usepackage{dutchcal}

\newcommand{\ee}{\end{equation}}
\newcommand{\bb}{\begin{equation}}
\newcommand{\eqb}{\begin{eqnarray}}
\newcommand{\eqf}{\end{eqnarray}}

\def\sigmavec{\mbox{\boldmath$\sigma$}}
\def\nablavec{\mbox{\boldmath$\nabla$}}

\begin{document}
\title{ {{Meissner Effect from Landau Problem}}}
\author{Paola Arias }
\email{paola.arias@usach.cl}
\affiliation{Departamento de  F\'{\i}sica, Universidad de  Santiago de
  Chile, Casilla 307, Santiago, Chile}
 \author{J. Gamboa }
\email{jorge.gamboa@usach.cl}
\affiliation{Departamento de  F\'{\i}sica, Universidad de  Santiago de
  Chile, Casilla 307, Santiago, Chile}
  \author{F. M\'endez }
\email{fernando.mendez@usach.cl}
\affiliation{Departamento de  F\'{\i}sica, Universidad de  Santiago de
  Chile, Casilla 307, Santiago, Chile}
   \author{David Valenzuela }
\affiliation{Departamento  de F\'{\i}sica,  Pontificia Universidad 
  Cat\'olica de 
  Chile, Casilla 306, Santiago 22, Chile}  
  
\date{\today}

\begin{abstract}
{{The Landau problem for inhomogeneous magnetic fields is examined in a very general context and several interesting analogies with the Nielsen-Olesen vortices are established. Firstly we show that the Landau problem with non-homogeneous magnetic fields exhibits Meissner effect that is unstable unless two-body interactions are added and vortices emerge. Using the scaling freedom we can write the Schr\"odinger equation in terms of  the scales ratio  $\kappa ={ E}/{ m }\propto 1- T/T_c
$ where the last identification is realised simply by using the Gizburg-Landau theory. We find our equations are valid in the superconducting regime, and it is not possible for the Cooper pairs amplitude to reach to a constant, non-zero value, and therefore the theory is unstable. The supersymmetric quantum mechanics version, by completeness, is also considered.}}
\end{abstract}
\date{\today}
\maketitle

\section{ Introduction}
The Meissner  effect is  a remarkable  experimental result  that shows
that the force  lines of an external magnetic field  are expelled when
penetrate a superconductor  sample \cite{sup}.  From a theoretical  point of view
this effect is explained by the  fact that the photons generate a mass
$m_\gamma  $ which  is the  inverse of  the penetration  depth in  the
superconductor.

Technically speaking, this effect is  obtained when the current $ {\bf
  J}$ -- which is  the source of Ampere's law -- acquires  a piece ${\bf
  J}  _L  = \alpha  {\bf  A}$  where  $\alpha$  is identified  {\it  a
  posteriori} with the mass of the photon  {and the relation between  
   $ {\bf J} _L$ and ${\bf A}$} is called the London equation.

The London equation  is an {\it ad hoc} relationship  that explains an
experimental  fact  but its  conceptual  content  is weak  unless  one
demonstrates that it comes from a fundamental explanation.

{On the other hand}, although the Ginzburg-Landau theory is made on purely intuitive basis,
its predictive  validity is  widely proven  not only  in the  field of
superconductivity  but also  in the  modern approach  of Bose-Einstein
condensation in statistical systems \cite{bose}, particle physics \cite{particles} and cosmology \cite{cosmos} .

{In spite of the fact that the} relationship  between superconductivity  and the  quantum
Hall  effect is  widely  discussed  in connection  with  anyons, {it seems that}  a
discussion in  between superconductivity and the  Landau problem \cite{witten}, {\it
  i.e.}  the motion of  charged particles in (non)homogeneous magnetic
fields, is less explored problem.

{This}  point of  view is very interesting, {we think},  because {it} helps to
clarify the relationship between the Ginzburg-Landau mean field theory \cite{rev} 
and the cornerstone Landau problem.
 
This  \lq\lq just   in  between"  region  is   also   interesting  because
{corresponds}  to the  transition  between  two non-perturbative  sectors
where,  in the first case, the role of Cooper pairs is dominant while in
the second case the presence of the external magnetic field instead of
the Cooper pairs becomes more important.
  
In  this paper  we  will  discuss the  connection  between the  points
outlined above and 
{we will} show
how in  the Landau problem  the Meissner  effect emerges, and  that
this  effect  would  be  destabilised   by  the  absence  of  two-body
interactions.
  
The stabilisation of the Meissner effect requires spontaneous symmetry
breaking, or in other words, { it requires}   the Hartree term
\[
H_{\mbox{\tiny{H}}}= \int dy~\psi^*(x) \psi^* (y) W(x-y) \psi (y) \psi (x).
\]
where {one   identifies}  $W(x-y)$,  the  two-body interaction,   with the  contact
term, {\it  i.e.} $W(x-y) =  \lambda \delta (x-y)$ with  $\lambda$ the
coupling constant.

In the absence of two-body interaction the spontaneous symmetry breaking does
not occur and the mass term for $\psi(x)$ has a sign ambiguity associated with one of the two phases 
of the system. These conclusions will be reached analytic and numerically.

The effects  due to  fermions {will be}  also {discussed}  using supersymmetric
quantum mechanics  techniques and {we will show that},  except for the  Pauli's term,
the results of the previous sections are maintained.

The paper is organised as follows: in section II we discuss the motion
of a charged particle in  no-homogeneous magnetic field and we discuss
the  Meissner effect  in  this context; in section  III  the role  of
fermions  is  discussed  by  using  supersymmetric  quantum  mechanics
techniques. {In the last section we discuss our results and we present  the 
conclusions.}

\section{Motion of charged particles in a homogenous magnetic field }

The strategy of the calculation is  similar to the motion of a charged
particle  in a  homogenous magnetic  field (Landau  problem). For purposes  of comparison  with standard results (particularly with
the Nielsen-Olesen vortex solution \cite{nielsen}) it is convenient to
describe the system with cylindrical coordinates ${\bf x}=\{\rho,\varphi,z\}$, 
and to take the vector potential as follows 
 \bb 
A_\rho = 0, \quad A_\varphi = \frac{\alpha  (\rho)}{\rho}, \quad A_z= 0. 
\label{cili}
\ee
The Hamiltonian describing a particle  of mass $m$ and electric charge
$q=+1$  in the  presence  of  a magnetic  field  ${\bf  B} =  (\alpha'
/\rho)\,\hat{z}$, turns out to be \footnote{Through all the text we use natural units.}
\eqb 
H  &=&   \frac{1}{2m}\left({\bf  p}-{\bf A} \right)^2   
\nonumber  
\\   
&=&   \frac{1}{2m}  \left(\hat{p}_\rho^2   + \frac{\hat{L}_z^2}{  \rho^2} + \hat{p}_z^2  -  
2\frac{\hat{L}_z A_\varphi}{\rho}  + A^2_\varphi \right) . 
\label{hamilt1} 
\eqf
with the operators $\hat{p}_z^2 =-\partial^2_z$,  $\hat{L}_z = -\imath\, \partial_\varphi$ and
$\hat{p}_\rho^2 =- \rho^{-1}\partial_\rho(\rho\partial_\rho)$. Since 
$z$ and $\varphi$ are cyclic variables, we take the following ansatz for the 
wave function $\psi (\rho,\varphi,z)$
\bb 
\psi (\rho,\varphi,z) =  e^{-iL_z \varphi} e^{ i p_z z}
f(\rho).
\label{ansatz}
\ee
where $p_z$ and $L_z$ are the conserved quantities corresponding to the 
cyclic  variables.  Will be useful to move to dimensionless variables, so let us define ${\bf x} \rightarrow {\bf x} \,m$, $A_\varphi\rightarrow A_\varphi/m$, ${\bf p}\rightarrow {\bf p}/m$ and $\psi\rightarrow \psi/m^{3/2}$. Correspondingly, the magnetic field has to be scaled as ${\bf B}\rightarrow {\bf B}/m^2 $ and the current source as ${\bf J}\rightarrow {\bf J}/m^3$.  Therefore, the Schr\"odinger equation becomes 
\bb
\left[ \hat{p}^2_\rho + \left( \frac{L_z}{\rho} - A_\varphi\right)^2+p_z^2\right]
f(\rho) = 2{\kappa^2} f(\rho),
\label{sch1}
\ee
where $\kappa^2=\frac{E}m$. 

Since the  wave function must be  single-valued for $\varphi\in[0,2\pi]$, then 
 $L_z  = n$  with $n \in {\mathbb Z}$. The Schr\"odinger equation reads
\bb
\frac{1}{\rho} \frac{\partial}{\partial \rho} \left( \rho
\frac{\partial f}{\partial \rho} \right)
+ \left[2\kappa^2- \left(\frac{n}{\rho} -A_\varphi \right)^2-p_z^2\right] f = 0.
\label{f1}
\ee
This equation can be solved, in principle, if  $A_\varphi(\rho)$ is given. In 
order to find this potential in a consistent way, we demand that the source of the 
magnetic field ${\bf B} =  (\alpha' /\rho)\,\hat{z}$ in the Ampere's law, to
be the conserved current for the Schr\"odinger equation (\ref{sch1}), that is,
the conserved current of probability
\bb
{\bf  J} =  \frac{1}{2i}\left(\psi^* \nablavec  \psi -\psi  \nablavec
\psi^*\right)  -{\bf A}\psi^* \psi. 
\label{current} 
\ee
For the ansatz  (\ref{ansatz}),  we find  
\bb
{\bf J} =  \left[\left(  \frac{L_z}{\rho} -A_\varphi \right) {f^2}\right]\, {\bf e}_{\varphi} +
{p_z^2}\,f^2\,{\bf e}_{z},
\ee
{with}  $\{{\bf e}_{\varphi} ,{\bf e}_{z}\}$ unit vectors. 
It is possible to choose -- without loss of generality -- a frame such that $p_z=0$. 
The  Ampere's law
\bb
\nablavec \times {\bf B} = {\bf J}, 
 \label{ampere}
\ee 
then, yields to 
\bb 
-\nablavec^2 {\bf A}_\varphi = \left( \frac{n}{\rho}  -A_\varphi \right) {f^2},
\ee
or
\begin{equation}
\label{ampe3}
\frac{1}{\rho}\frac{\partial}{\partial \rho}\left(\rho \frac{\partial\,A_\varphi}{\partial\rho}\right) -\frac{A_\varphi}{\rho^2} +
\left( \frac{n}{\rho}  -A_\varphi \right){f^2}=0
\end{equation}
We can write now the set of equations  for $f$ and $A_\varphi$ (namely, the Schr\"odinger equation and the 
Ampere's law) in terms of $\alpha$, using the ansatz of (\ref{cili}) and we find
 \begin{eqnarray}
 \label{schofin}
 f'' + \frac{f'}{\rho}  +\left[2 {\kappa^2}-\frac{1}{\rho^2}(n-\alpha)^2\right]f &=& 0,
 \\
 \label{alpha1}
 \alpha''    - \frac{\alpha'}{\rho}    +   \left(  n-   \alpha  \right)
f^2&=&0.
 \end{eqnarray}
This   set of  equations  describes the motion of our system.  In order
to discuss the solutions of this set, we will consider the case $n=1$ in the 
rest of the text. 

Before moving on to the analysis and solving of these coupled equations, a comment about the physics of this model is in order. From equation~(\ref{sch1}), with $p_z=0$, we can find the energy functional that leads to such equation. We call $\mathcal E$ the (dimensionless) total energy per unit of length of the system, and we have
\bb
\mathcal E=\int  d^2x\left[ \frac{1}{2} \left(\nabla \times {\bf A}\right)^2+\frac{1}2| \left(-i \nabla-{\bf A}\right) \psi|^2-\kappa^2 |\psi|^2\right].
\ee
Using the chosen vector potential given in~(\ref{cili}), it can be easily checked that the above functional gives both, Schr\"odinger and Ampere's law. We can compare this energy with the one of a superconducting sample, which is usually written as \cite{Kleinert:1989kx}
\eqb
\mathcal E&=&\int  d^2x\left[ \frac{1}{2} \left(\nabla \times {\bf A}\right)^2+\frac{1}2| \left(-i \nabla-{\bf A}\right) \psi|^2+\right. \nonumber \\
&&\left.\frac{a_0}2\left(\frac{T}{T_c}-1\right) |\psi|^2+\frac{g}{4}|\psi|^4\right]
\eqf
The $|\psi|^4$ term gives {the} spontaneous symmetry breaking that makes possible the second order phase transition, and $a_0$ and $g$ are dimensionless constants. The important point for us is that we can make the following {identification}
\bb
\kappa^2=\frac{a_0}2\left(1-\frac{T}{T_c}\right),
\ee
which implies that our equations are valid in the superconducting phase, meaning $T\leq T_c$. We note that $\kappa \ll1$ should represent a sample in the vicinity of the phase transition point, whereas a greater $\kappa$ corresponds to one deep into the superconducting regime.

Coming back to our set of equations (\ref{schofin}) and (\ref{alpha1}), {they} can be solved
for a set of  boundary conditions. Such conditions will be obtained from 
 physical requirements in the present approach.

The first requirement refers to the magnetic field.  We demand the following 
values at  boundaries (in the rescaled dimensionless variables)
\begin{equation}
\label{bboundary}
{B} = \left \{
\begin{array}{ll}
 1 &\quad  \mbox{for}~~ \rho\to 0,
\\
 0 &\quad \mbox{for} ~~\rho\to \infty,
\end{array} \right.
\end{equation}
with $B = |{\bf B}|$ and $
{\bf  B} =  \nablavec \times  {\bf A} =    \frac{\alpha'}{\rho}\,\hat{z}.$

Equation (\ref{bboundary}) imposes Meissner effect and for $\alpha$ it is 
equivalent to 
\begin{equation}
\label{bboundary2}
\alpha(\rho) = \left \{
\begin{array}{ll}
 \frac{\rho^2}{2} &\quad  \mbox{for}~~ \rho\ll 1,
\\
 & 
 \\
 1 &\quad \mbox{for} ~\rho\gg 1.
\end{array} \right.
\end{equation}
In particular, this behaviour sets the following initial conditions $\alpha (0) =0$, and $\alpha'(0)=1$ or,
equivalently, the following boundary conditions $\alpha(0) =0$ and $\alpha(\infty) = 1$, where the 
symbol \lq$\infty$\rq ~ stands for $\rho \gg1$.

The boundary conditions for $f $ are determined by our second 
requirement that current is finite in the near region $\rho\approx 0$, while it 
should vanish in the far region $\rho \gg 1$. From (\ref{current}), it is direct
to check that both conditions hold if 
\begin{equation}
\label{boundaryf}
f(\rho) = \left \{
\begin{array}{ll}
 0 &\quad  \mbox{for}~~ \rho \ll 1,
\\
 & 
 \\
 \mbox{constant}  & \quad \mbox{for} ~\rho\gg 1.
\end{array} \right.
\end{equation}
\subsection{Asymptotic behaviour of the fields}
\begin{figure}
\begin{center}
\includegraphics[scale=0.7]{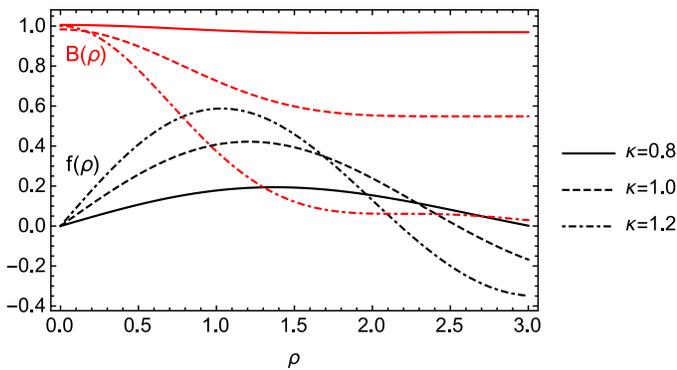}
\caption{\small{Magnetic field (red) and Cooper pair amplitude (black) as a function of the distance (in dimensionless units) for different values of $\kappa$. For smaller $\kappa$ we expect to be at the beginning of the superconducting phase, $T\sim T_c$, and for bigger $\kappa$, at temperatures below the phase transition $T<T_c$.  }}
\label{fig:meis}
\end{center}
\end{figure}

In order to see if the assumed asymptotic behaviour of $\alpha$ and $f$ given in ~(\ref{bboundary2}) and~(\ref{boundaryf}) are consistent with our set of equations (\ref{schofin})-(\ref{alpha1}), let us perform an analysis in both limits.
\begin{itemize}
\item[a)] for $\rho\ll1$, we can decouple the set of equations, considering $\alpha(\rho)\ll1$ and $f(\rho)\ll1$ and we obtain
\eqb
f''+\frac{f'}\rho-\frac{f}{\rho^2}=0,
\\
\alpha''-\frac{\alpha'}r=0,
\eqf
which accounts for a behaviour of $\alpha(\rho)\sim c_1 \left(\frac{\rho}{\kappa}\right)^2$ and $f(\rho)\sim c_2\frac{\rho}{\kappa}$. Where $c_1, c_2$ are constants.

\item[b)] For $\rho\gg1$ we consider $\alpha(\rho)\sim 1+\epsilon_1(\rho)$, with $\epsilon\ll1$, in order to have attenuation of the magnetic field at greater distances (Meissner effect), therefore the equation~(\ref{alpha1}) gets replaced by
\bb
\epsilon_1''-\frac{\epsilon_1'}{\rho}-\epsilon_1 f^2=0.
\ee
We see that if $f\approx $ const in the last equation, we obtain for $\alpha(\rho)\sim 1+\rho K_1( f\rho )$, however, the replacement of this expression back into eq.~(\ref{schofin}), for $f\sim $ const,  shows that $f\approx 0$. 
Thus, we have to instead consider the following asymptotic equation
\bb
\epsilon_1''-\frac{\epsilon_1'}{\rho}=0,
\ee
whose solutions are $\alpha\sim 1+b_1+ b_2 \rho^2$, with $b_1,b_2$ constants, which is consistent with the Meissner effect for $b_2\rightarrow 0$. Therefore, the asymptotic behaviour of $f(\rho)$ that is consistent with expulsion of the magnetic field in the sample is $f(\rho)\rightarrow 0$ when $\rho\rightarrow \infty$. 
\end{itemize}
In order to find numerical solutions for our sets of differential equations we implement a shooting method to match the solution near $\rho \rightarrow 0$ and $\rho\rightarrow \infty$. In order to do so, we consider for $\rho\ll1$ a polynomial behaviour $f$ and $\alpha$ and for $\rho\gg1$ we can linearise the equations as we did above, considering $\alpha(\rho)\sim 1+\epsilon_1(\rho)$ and $f(\rho)\sim \epsilon_2(\rho)\ll1$. Thus,
\eqb
\epsilon_2''+\frac{\epsilon'_2}\rho+2{\kappa^2}\epsilon_2=0,
\\
\epsilon_1''-\frac{\epsilon_1'}{\rho}=0,
\eqf 
with solutions
\eqb
\alpha(\rho)&=&1+b_1+b_2 \rho,
\\
f(\rho)&=&d_1 J_0({\sqrt{2}\kappa\,\rho})+d_2Y_0({\sqrt{2}\kappa\,\rho}),
\eqf
where $J_0$ and $Y_0$ are Bessel functions and $b_1, b_2, d_1$ and $d_2$ are constants of integration.
In figure~(\ref{fig:meis}) we show the magnetic field and Cooper pair amplitude as a function of the distance for several values of $\kappa$. For smaller $\kappa$ the attenuation of the magnetic field is almost negligible, and also the growing of the Cooper pair amplitude, this corresponds to a temperature very near the phase transition, $T=T_c$. As $\kappa$ gets larger (and therefore the temperature goes much below $T_c$) the effects on the magnetic field are more significant. We can understand this behaviour since, for $\rho<1$, the missing $f^3$ term that appears in the usual Ginzburg-Landau theory can be neglected in comparison to the linear term \cite{abrikosov}. Thus, near the border of the superconductor, our model is a good approximation of the superconductivity theory. But as $\rho>1$, the cubic term is needed to stabilise the Meissner effect. This last statement can be used to understand why the amplitude of the Cooper pairs fails to reach a constant asymptotic behaviour, since the contact term between electrons is missing.  This translates to a magnetic flux inside the sample which is not quantised and therefore, no vortices can appear using this theory. 

\section{The role of fermions}
In a conductive material the  carriers are electrons and therefore one
should formulate  the above problem including  the fermionic character
of the carriers.   From the Hamiltonian point of view  this means that
we must put from the beginning the Pauli term $\sigma_{\mu \nu} F^{\mu
  \nu}$  that  for  the  static  magnetic field  this  is  reduced  to
$\sigmavec. {\bf B}$. 

In order to  explain the previous statement we  will proceed following
an approach  based in supersymmetry  proposed by  one of us  long time
ago \cite{jz} \footnote{ The literature in supersymmetric quantum mechanics is very extensive and in the context of our discussion  see  \cite{gozzi}, \cite{witten1} and \cite{coop}.}.  Thus,  we  start  by considering  the  following  supersymmetric
charges 
\eqb 
S &=&  \left( D_i  + \partial_i W  \right) \sigma_i  \otimes \sigma_-,
\nonumber 
\\
S^{\dagger} &=&  \left( -D_i +  \partial_i W \right)  \sigma_i \otimes
\sigma_+, \label{super1} 
\eqf
where  $\sigma_{\pm}=  \sigma_1 \pm  i  \sigma_2$  and $\sigma_i$  are
Pauli's matrices, $D_i$  is the covariant derivative  defined as $D_i=
\partial_i -A_i$, {W is the superpotential}  and 
\[
\left[D_i,D_j\right] = -i F_{ij},
\] with $B_k= \frac{1}{2} \epsilon_{kij}F_{ij}$,{ the} magnetic field.

In addition, supercharges must satisfy the algebra
\eqb 
\{ S,S^{\dagger} \} &=& 2 H_s, \label{hamss}
\\
\left[ H_s,S\right] &=&0=\left[ H_s,S^{\dagger}\right], 
\\
S^2 &=& 0= {S^{\dagger}}^2.
\eqf

The explicit  calculation of  (\ref{hamss}) defines  $H_s$ an  then, by
using (\ref{super1}),  we find 
\bb 
H_s  =  \frac{1}{2}  \left({\bf  p} -{\bf  A}\right)^2  -  \frac{1}{2}
\sigmavec \cdot {\bf B} + \left( \nablavec W\right)^2 -\nablavec^2 W ~\sigma_3,
\ee
with $\sigma_3 = \left(\begin{array}{cc} 1 &0
\\
0 & -1\end{array} \right)$.

The  last   term {in the previous equation is identified 
with the potential energy through}  $\left(  \nablavec  W\right)^2   \mp  \nablavec^2  W
=u_{\mp}({\bf x})$, 
{defining    a Ricatti equation for W}.  
 The signs $\pm$ are related to  the normalization of the ground state
 and, for example,  $\psi_0^{(+)}$ means that the spinor {that corresponds  to the ground 
 state} is chosen as 
 \[
 \psi_0^{(+)} = \left(\begin{array}{c} e^{-W} 
 \\ 0\end{array}\right).
 \]
The ground state is  {then}  related to $W$ through $W = - \ln  \psi $ and $W$
is the  superpotential.  So, since  the Hamiltonian is hermitian,  $ W
\in {\Re }$, then in symbolic form we can write the Hamiltonian as
\bb 
H^{\pm}_s = \frac{1}{2} \left({\bf  p} -{\bf A}\right)^2 - \frac{1}{2}
\sigmavec \cdot   {\bf B}  + \left(  \nablavec W\right)^2  \mp \nablavec^2
W, \label{ssh} 
\ee 
understanding, of  course, that  one and only  one component  of the
spinor is normalizable. 

If we choose $  {\bf B} = (0,0, B_3 (x))$, which is  what we have been
considering, then 
\bb
H^{\pm}  =  \frac{1}{2}  \left({\bf  p}  -{\bf  A}\right)^2  +  \left(
\nablavec  W  \right)^2  \mp  \left( \nablavec^2  W  -\frac{1}{2}  B_3
(x)\right). \label{fermion1} 
\ee
this is the fermionic extension of the Hamiltonian (\ref{hamilt1}). 

\section{Discussion and Conclusions}

The motion of charged particles in an external (constant) magnetic field  is one of the most interesting problems of contemporary physics, 
{and is present in  } 
many other topics such as anyons, cosmic strings, Aharonov-Bohm effect, cosmology, quantum Hall effect and so on. 

The mechanism itself is 
remarkable because it makes use of the gauge invariance (and its topological implications) through the modification of the canonical commutators thus providing a natural link with highly sophisticated mathematics (as for example  noncommutative geometry and Poissonian manifolds). 

However, we think that the Landau problem, properly modified, can also be useful to understand the physical basis that connects the superconductivity and the quantum Hall effect. 

The results that we have presented in this paper are a step towards this direction. We have shown how to approach the critical points on both sides in a phase transition region and we have shown that even not being in the critical phase equally one can find an interesting phenomenon {resembling the 
Meissner effect}

This work  was  supported  Dicyt and USA-1555 (J.G.) and  Fondecyt-Chile project 
 1161150 (P.A.)

\end{document}